\documentclass[12pt]{article}
\usepackage{amsfonts,amsmath}

\newcommand{\next}{\\[-5.4mm]\pagebreak[1]}

\newtheorem{thm}{Theorem}

\newtheorem{lem}[thm]{Lemma}
\newtheorem{prop}[thm]{Proposition}
\newtheorem{defn}{Definition}

\newenvironment{proof}{
\par
\noindent {\bf Proof.}\rm}%
{\hspace*{\fill}\rule{0.5em}{0.809em}\par}

\numberwithin{equation}{section}
\newcommand{\Prs}{{\textstyle \Pr_s}}
\newcommand{\pbfrac}[2]{\mbox{$\mbox{}^{#1}\!/_{#2}$}}
\newcommand{\neopbfrac}[2]{\mbox{$\mbox{}^{#1}\!/_{\mbox{\footnotesize{$#2$}}}$}}

\newcommand{\Z}{\phantom{-}}
\newcommand{\ii}{\imath}
\def\bra#1{\langle#1|}
\def\ket#1{|#1\rangle}
\newcommand{\phiplusn}{\frac{1}{\sqrt{2}}
\ket{0^n}+\frac{1}{\sqrt{2}}\ket{1^n}}
\newcommand{\phiminusn}{\frac{1}{\sqrt{2}}
\ket{0^n}-\frac{1}{\sqrt{2}}\ket{1^n}}
\newcommand{\optprob}{\frac{1}{2} + 2^{-\ceil{n/2}}}
\def\ceil#1{\lceil#1\rceil}
\title{Quantum Pseudo-Telepathy}

\author{\large Gilles Brassard\,%
\thanks{\,Supported in part by Canada's {\sc Nserc}, Qu\'ebec's {\sc Fqrnt},
the Canada Research Chair \mbox{programme},
the Canadian Institute for Advanced Research (CIAR), the
Mathematics of Information Technology and Complex Systems Network (MITACS)
and the Canadian Institute for Photonic Innovations (CIPI).}
~~~~~Anne Broadbent\,%
\thanks{\,Supported in part by a scholarship from Canada's {\sc Nserc}.}
~~~~~Alain Tapp\,%
\thanks{\,Supported in part by Canada's {\sc Nserc}, Qu\'ebec's {\sc Fqrnt},
the CIAR and MITACS.}
\\ {\normalsize\it D\'epartement~IRO, Universit\'e de Montr\'eal}\\[-1ex]
{\normalsize\it C.P.~6128, succursale centre-ville}\\[-1ex]
{\normalsize\it Montr\'eal (Qu\'ebec),  H3C~3J7 \textsc{Canada}}\\
{\normalsize\texttt{\{brassard,\,broadbea,\,tappa\}}\textbf{\char"40}\texttt{iro.umontreal.ca}}}

\begin{document}
\maketitle
\thispagestyle{empty}

\begin{abstract}
Quantum information processing is at the crossroads of physics,
mathematics and computer science. It is concerned with that we can
and \mbox{cannot} do with quantum information that goes beyond the abilities
of \mbox{classical} information processing devices.  Communication complexity
is an area of classical computer science that aims at quantifying the amount
of communication necessary to solve distributed computational problems.
\emph{Quantum} communication complexity uses quantum mechanics to
reduce the amount of communication that would be classically required.

\emph{Pseudo-telepathy} is a surprising application of quantum
information \mbox{processing} to communication complexity.  Thanks to
entanglement, perhaps the most nonclassical manifestation of quantum
mechanics, two or more quantum players can accomplish a distributed task
with \emph{no} need for communication whatsoever,
which would be an impossible feat for classical players.

After a detailed overview of the principle and purpose of
pseudo-telepathy, we present a survey of recent and no-so-recent work on
the subject.  In~particular, we describe and analyse all the
pseudo-telepathy games currently known to the authors.
\end{abstract}

\vspace{5mm}
\section{Introduction}

\subsection{Quantum Nonlocality}

Albert Einstein was convinced that the physical world is local and realistic.
By~\textit{local}, he meant that no action performed at some location
\textsf{A} could have an effect at some other remote location \textsf{B}
in a time faster than that required by light to travel from \textsf{A}
to~\textsf{B}\@.   In~particular, instantaneous \mbox{action} at a distance
would be impossible. By~\textit{realistic}, he meant that measurements
could only reveal ``elements of reality'' that were already present
in the system being measured.
He~attributed Heisenberg's uncertainty relations
to the fact that any measurement of a particle's position,
no matter how subtle, would disturb its momentum, and vice versa,
and therefore one could never know both position and momentum
simultaneously with arbitrary precision.

But Niels Bohr said no. In his interpretation of quantum mechanics,
Heisenberg's uncertainty relations are unavoidable because it is
impossible for a particle to have both its position and momentum
simultaneously \textit{defined} with arbitrary precision.
It is the act of
measuring one or the other that makes \textit{it} come to existence.
How could you disturb that which does not even exist in the first place?
You can't deny Einstein's realism more strongly than this!

In an attempt to ``prove'' his point, Einstein wrote a famous paper in 1935
with Boris Podolsky and Nathan Rosen~\cite{EPR35}. In~that paper, they
introduced
the notion of quantum \textit{entanglement}, according to which it would be
possible in some cases to determine the position \textit{or} the
momentum of a particle without in any way interacting with it.
Based on the locality assumption,
they concluded that \textit{both} position and momentum had to be
simultaneously well-defined for that particle, contradicting
Bohr's view.

In a curious turn of history, John Bell proved three decades later that
the entanglement phenomenon predicted by Einstein, Podolsky and Rosen
entails statistics that would be impossible to explain by any
local realistic theory whatsoever~\cite{Bell}.
In~other words, Einstein's argument had backfired, with entanglement
becoming the most convincing argument in favour of Bohr's view!
Despite appearances, quantum mechanics does not really contradict the
locality assumption of Einstein: It~is still the case that no action performed
at~\textsf{A} can have an instantaneous \textit{observable} effect
at~\textsf{B}\@. But it certainly does contradict realism.

It remained to perform actual experiments, which was done in the 1980s
by Alain Aspect \textit{et~al}, and sure enough the predictions of quantum mechanics come
true~\cite{Aspect1,Aspect2,Aspect3}.  \mbox{Subsequent} experiments were conclusively
conducted by Nicolas Gisin and collaborators over more than 10~kilometres
of distance between points \textsf{A} and~\textsf{B}~\cite{gisin98}, and
even in relativistic settings in which both
detectors, each in its own inertial reference frame,
are first to do the measurement~\cite{gisin01}!
To~get a convincing experimental
demonstration that the physical world is not local realistic, we must
make a large number of careful experiments and collect significant
statistical evidence that such data would be overwhelmingly unlikely
in a classical local realistic world.
Moreover, the apparatus must be sufficiently accurate to rule out
a variety of loopholes that would make a classical ``explanation''
possible again---more on this issue in Sections~\ref{convincing}
and~\ref{loopholes}.

Despite David Mermin's heroic effort at explaining Bell's
theorem~\cite{Mermin81b},
it is not so easy to convey the significance of these
experiments to a non-scientist because nontrivial probabilistic arguments
are involved. It would be much nicer to exhibit an observable behaviour
that is \textbf{obviously} impossible in any classical world so
dear to Einstein\,\footnote{\,For a quantum physicist, even relativity
belongs to classical physics.}.
A~significant step in the right direction was taken by
Daniel Greenberger, Michael Horne and Anton Zeilinger~\cite{GHZ},
and later by
Lucien Hardy~\cite{hardy}, when they introduced quantum phenomena
that can be shown to be classically impossible without resorting
to probabilities: A~classical contradiction results from what is
and what is not possible in the observed behaviour, not from the
specific probabilities of various events.
These phenomena have been called by such names as
``Bell's theorem without inequalities''.
It~is our belief that pseudo-telepathy---which is the topic of this
survey---takes another step in this direction, even though the
Greenberger-Horne-Zeilinger conundrum (but \textit{not} Hardy's)
can be interpreted as an early instance of pseudo-telepathy.

It is important to mention a common misconception in reasoning about
experimental realizations of Bell's theorem without inequalities.
Too often, we read that they allow to rule out local realism
in  a \emph{single run}: ``\,The quantum nonlocality can
thus, in principle, be manifest in a single run of a certain
measurement''~\cite{CPZBZ03}. The fallacy here is that if the
players (classical or quantum) win for a single run, we cannot
conclude anything: they could have been lucky at guessing.
As~Asher Peres
wrote, about those who made this mistake: ``The list of authors is
too long to give explicitly, and it would be unfair to give only a
partial list''~\cite{Peres2000}.

\subsection{Pseudo-Telepathy}

Consider two parties, whom we shall call Alice and Bob.
They pretend to be endowed with telepathic powers.
To~convince scientists Xavier and Yolande, they agree to play
the following game.
First, Xavier and Yolande secretly decide on two lists of animals,
such as that shown in Table~\ref{animals}.
\begin{table}
\caption{Example of a game.}
\label{animals}
\begin{center}
\begin{tabular}{c|c|c}
Round & \textsf{Xavier} & \textsf{Yolande} \\ \hline
1 & lion & {\huge \phantom{I}}giraffe{\LARGE \phantom{I}} \\
2 & tiger & tiger \\
3 & hyena & elephant \\
4 & crocodile & alligator \\
5 & platypus & platypus \\
$\cdots$ & $\cdots$ & $\cdots$
\end{tabular}
\end{center}
\end{table}
Then, Xavier takes Alice far away from Bob and Yolande.
At predetermined times, Xavier and Yolande name animals from their
lists to Alice and Bob. For example, Xavier and Yolande simultaneously
say ``lion'' and ``giraffe'', respectively.  Without consulting each other,
Alice and Bob must immediately decide whether or not they were presented
with the same animal; in this case, they should both answer ``no!''.
If~Alice and Bob succeed systematically in a sufficiently long
sequence of trials, Xavier and Yolande will conclude
that Alice and Bob are able to communicate somehow.
But if communication is classically impossible because Alice and
Bob are sufficiently far apart that a signal from Alice going at the
speed of light would not reach Bob in time to influence his answer
(and vice versa from Bob to Alice), then Xavier and Yolande would
be forced into believing that Alice and Bob are able to communicate
in a way unknown to (classical) physics.  In~this case, telepathy
would not seem to be worse than any other esoteric ``explanation'',
would~it?

It turns out that quantum mechanics cannot help Alice and Bob win
this animal-guessing game.  (Otherwise, a solution to this game
could be harnessed to provide faster-than-light signalling.)
But there are other similar games that
are equally impossible for classical players, yet they
can be won systematically, without any communication, provided
Alice and Bob share prior entanglement.  This is the phenomenon
we call ``pseudo-telepathy'' because it would appear as magical
as ``true'' telepathy to a classical physicist, yet it has a
perfectly scientific explanation: quantum mechanics.

There are several reasons to be interested in pseudo-telepathy games.
\begin{itemize}
\item[$\Diamond$] Some of these games are conceptually very simple.
Their classical impossibility can be explained in a few minutes
to any reasonably intelligent person.
This makes their quantum possibility all the more impressive.
The best examples in this category are the Mermin-GHZ game
and the magic square game, described in Sections~\ref{parity}
and~\ref{magic}, respectively.

\item[$\Diamond$] If implemented successfully, any one of these
games would provide
\mbox{extremely} convincing evidence that classical physics does not rule
the world for anyone who is already convinced that faster-than-light
signalling is \mbox{impossible} (and therefore that it is \textit{not}
because Alice and Bob are able to communicate, however covertly, that they
win the game systematically).

\item[$\Diamond$] The theoretical and experimental study of
pseudo-telepathy has the potential to provide the first loophole-free
demonstration that the physical world cannot be local realistic.

\item[$\Diamond$] In introducing the broader field of quantum communication
complexity in~1993, Andrew Yao asked if quantum mechanics could
be used to reduce the amount of communication required to solve some
distributed computational tasks, but he could not answer the
question~\cite{Yao93}.
Surprisingly, a pseudo-telepathic version of this
question had already been solved ten years previously~\cite{HR82}
by Peter Heywood and Michael Redhead\,\footnote{%
\,One could argue that pseudo-telepathy games do \textit{not}
solve Yao's problem as stated and therefore the first solution to
Yao's problem came as late as 1997.}---see
Section~\ref{section:impossiblecoulouring}.
Please consult~\cite{QCC} for a survey of quantum communication
complexity and its interplay with pseudo-telepathy.

\end{itemize}

We are now ready to give a formal definition of pseudo-telepathy.
For the sake of simplicity, we define it in the context of two
players, Alice and Bob.  The generalization to multi-party
settings is obvious.

A two-party game is defined as a sextuple
\mbox{$G=\langle X, Y, A, B, P, W \rangle$},
where $X$, $Y$, $A$ and $B$ are sets,
\mbox{$P \subseteq X \times Y$} and
\mbox{$W \subseteq X \times Y \times A \times B$}.
It~is convenient to think of $X$ and $Y$ as the \textit{input} sets,
$A$ and $B$ as the \textit{output} sets, $P$ as a predicate on
\mbox{$X \times Y$} known as the \textit{promise},
and $W$ as the \textit{winning condition}, which is a relation
between inputs and outputs that has to be satisfied by Alice and Bob
whenever the promise is fulfilled.
[In~the generalization to 3 parties, we use $X$, $Y$, $Z$ as input
sets and $A$, $B$, $C$ (for Charlie) as output sets; $P$ and $W$ are changed
accordingly. In~the \mbox{$n$-party} generalization, we use
\mbox{$X = X_1 \times X_2 \times \cdots \times X_n$} and
\mbox{$A = A_1 \times A_2 \times \cdots \times A_n$} as input
and output sets, respectively,
$P$~becomes a predicate on~$X$ and $W$ becomes
a relation on \mbox{$X \times A$}.]

At the outset of the game, Alice and Bob are allowed to discuss strategy
and exchange any amount of classical information, including the value
of random variables.  This is carried out in secret of Xavier and Yolande.
If~Alice and Bob are quantum players, they are
also allowed to share unlimited amounts of entanglement.
Xavier and Yolande are also allowed to discuss strategy in secret
of Alice and Bob.
Then, Alice and Bob are separated and they will not be allowed
to communicate any further until the game is over.
In~each \textit{round} of the game, Alice and Bob are presented
by Xavier and Yolande with inputs \mbox{$x \in X$} and \mbox{$y \in Y$},
respectively.
The task of Alice and Bob is to produce outputs
\mbox{$a \in A$} and \mbox{$b \in B$}, respectively.
We~say that Alice and Bob \textit{win the round} if
either \mbox{$(x,y) \not\in P$} (the promise does not hold) or
\mbox{$(x,y,a,b) \in W$}.
The pairs \mbox{$(x,y)$} and \mbox{$(a,b)$} are called
the \textit{question} and the \textit{answer} for this round, respectively.
Any question \mbox{$(x,y)$} is said to be \textit{legitimate}
if it belongs to~$P$,
and any answer \mbox{$(a,b)$} is said to be \textit{appropriate}
if \mbox{$(x,y,a,b) \in W$}.
We~say that Alice and Bob \textit{win the game} if
they keep winning round after round.
Finally, we~say that Alice and Bob \textit{have a winning strategy}
if they are mathematically certain to win the game as long as they
have not exhausted all the classical information and quantum entanglement
(if any) shared at the outset of the game.
Please note the difference between winning the game and
having a winning strategy: no matter how many rounds Alice and Bob
have won, only \textit{statistical evidence} can be gathered by
Xavier and Yolande towards
the hypothesis that they indeed have a winning strategy,
but a single lost round suffices to discredit any claim to
a winning strategy.  (See~Section~\ref{convincing} for
practical considerations.)

To continue our earlier example from Table~\ref{animals},
$X$ and $Y$ would be nontrivial sets of animal names,
\mbox{$A=B=\{\textsf{yes},\textsf{no}\}$},
\mbox{$P=X \times Y$} (which is to say that all questions are legitimate)
and \mbox{$(x,y,a,b) \in W$} if and only if either \mbox{$x=y$}
and \mbox{$a=b=\textsf{yes}$} or \mbox{$x \neq y$}
and \mbox{$a=b=\textsf{no}$}.
In~the first round, for instance, Xavier and Yolande ask the question
\mbox{(lion,\,giraffe)} and Alice and Bob should provide
the answer \mbox{(\textsf{no},\,\textsf{no})}.
As~we said before, this game does not admit a winning strategy,
be it classical or quantum,
but Alice and Bob could still win by being exponentially lucky
in the number of rounds.

Finally, we say that a game exhibits \textit{pseudo-telepathy}
if it does not admit a winning strategy whenever Alice and Bob
are restricted to being classical players, yet it does admit
a winning strategy provided quantum players Alice and Bob share
a sufficient supply of entanglement.

We can quantify the quantumness of a pseudo-telepathy
game by how bad classical players would be at~it.  This is interesting
because the harder a pseudo-telepathy game is for classical players,
the more convincing it is when quantum players win it round after round.
A~more subtle reason to be interested in this issue is that quantum
players will not be perfect in real life: it is unavoidable that they will
lose a few rounds now and then.  But if they are still better at the game
than any classical player could be, the main purpose of proving that the
world cannot be local realistic will still be fulfilled.
This is how we hope to use pseudo-telepathy games to circumvent
the various loopholes that have plagued until now every attempt at
reaching this holy Grail~\cite{GZ99,Massar02,BCHKP02}.

A classical strategy is \textit{deterministic} if there are two functions
\mbox{$f:X \to A$} and \mbox{$g:Y \to B$} such that Alice and Bob
systematically output $f(x)$ and $g(y)$ on inputs $x$ and $y$,
respectively.  The success of a deterministic strategy is defined
by the \textit{proportion} of legitimate questions on which it provides
an appropriate answer.
Given a game $G$, we let $\widetilde{\omega}(G)$ stand for the
maximum success proportion, over all possible deterministic
strategies, for classical players that play game~$G$\@.  Formally:
\begin{equation}
 \widetilde{\omega}(G) = \max_{f,g}\frac{\#\{(x,y) \in P \mid
          (x,y,f(x),g(y)) \in W \}}{\#P}
\end{equation}

Classical players can implement \textit{probabilistic} strategies when
they are allowed to toss local coins and share random variables.
In~that case, we are interested in the \textit{probability} that the
strategy succeeds on given questions.  We shall judge the merit of
a probabilistic strategy by its performance on the worst possible
legitimate question.  More precisely, we say that a probabilistic strategy
is \textit{successful with probability~$p$} if it produces an appropriate
answer with probability at least $p$ on all legitimate questions.
Note that any deterministic strategy that does not win the game
with certainty is ``successful'' with probability zero since there
is at least one legitimate question on which it fails systematically.
Given a game $G$, we let ${\omega}(G)$ stand for the
maximum success probability, over all possible probabilistic
strategies, for classical players that play game~$G$\@.
Formally, given a strategy $s$ and a legitimate question \mbox{$(x,y)$},
let \mbox{$\Prs(win \mid (x,y))$} denote the probability that strategy~$s$
provides an appropriate answer on that question.  Then
\begin{equation}
 \omega(G) = \max_s \min_{(x,y) \in P} \Prs(win \mid (x,y))
\end{equation}

The next theorem provides a simple yet important relation between the
maximum success proportion for deterministic strategies and the
maximum probability of success for probabilistic strategies.
But first we need a technical proposition.

\begin{prop}
\label{prop:proportion}
Let $G$ be a game.  Consider any probabilistic strategy~$s$.
If~the questions are asked uniformly at random among all legitimate
questions, the probability that the players win using~$s$ is
$\widetilde{\omega}(G)$ at best.
\end{prop}

\newpage

\begin{proof}
We consider a general probabilistic strategy~$s$, which is a
probability distribution over a finite set of deterministic
strategies, say \mbox{$\{s_1, s_2, \ldots s_m\}$}.  Let $\Pr(s_i)$ be the
probability that strategy $s_i$ be chosen, and let
\mbox{$p_i \le \widetilde{\omega}(G)$} be the
success proportion of strategy~$s_i$.  The probability that the
players win the game according to strategy~$s$ is:
\begin{equation}
\sum_{i=1}^m p_i \Pr(s_i)
~\leq~ \widetilde{\omega}(G) \sum_{i=1}^m \Pr(s_i) \\
~=~ \widetilde{\omega}(G)
\end{equation}
\end{proof}

\begin{thm}\label{framework:probabilistic}
For any game $G$,
\mbox{${\omega}(G) \le \widetilde{\omega}(G)$}.
\end{thm}
\begin{proof}
Consider any probabilistic strategy $s$ that is successful with maximal
probability $\omega(G)$. By~definition of $\omega(G)$, given any
legitimate question \mbox{$(x,y)$},
\mbox{$\Prs(win \mid (x,y)) \geq \omega(G)$}. If~the question is chosen
uniformly at random, the probability $q$ of winning the game using
strategy $s$ is
\begin{equation}
q ~= \sum_{(x,y) \in P} \frac{\Prs(win \mid (x,y))}{\#P}
~\geq \sum_{(x,y) \in P} \frac{\omega(G)}{\#P}
= \omega(G)
\end{equation}
By proposition~\ref{prop:proportion}, $\widetilde{\omega}(G) \geq q$,
and since \mbox{$q \geq \omega(G)$}, then
\mbox{$\widetilde{\omega}(G) \geq \omega(G)$}.
\end{proof}

The next observation is useful to determine values and bounds on
$\widetilde{\omega}(G)$ and~${\omega}(G)$.
\begin{lem}
\label{lemma:proportion}
Let $G$ be a game whose set of legitimate questions is~$P$.
If~it does not admit a classical winning strategy, then
\begin{equation}
 \omega(G) \leq \widetilde{\omega}(G) \leq \frac{\#P-1}{\#P}
\end{equation}
\end{lem}
\begin{proof}
Since $\widetilde{\omega}(G)$ is the maximum success
proportion, over all possible deterministic strategies, for
classical players that play game~$G$, it is the ratio of
some integer to
the total number $P$ of legitimate questions.
Since the game does not admit a classical winning strategy,
this ratio is not $1$.  The
next best alternative, which may or may not be achievable by some strategy,
would be that \mbox{$\widetilde{\omega}(G) = (\#P-1)/\#P$}.
The~Lemma follows from Theorem~\ref{framework:probabilistic}.
\end{proof}

\subsection{How Convincing Can Actual Experiments Be?}\label{convincing}
A major motivation for studying pseudo-telepathy games is that their
physical implementation could provide increasingly convincing and
loophole-free demonstrations that the physical world is not
local realistic. \mbox{After} estab\-lishing that a given
game is indeed a pseudo-telepathy game,
the ideal experiment would consist in implementing the
quantum winning strategy and running it on a significant number of rounds
until either:
\begin{itemize}
\item[$\Diamond$]
the players lose even a single round,
in which case the predictions of quantum mechanics
have been shown to fail, demonstrating that quantum \mbox{mechanics}
must be wrong, and it's back to the drawing board; or
\item[$\Diamond$]
the players win systematically enough rounds
to rule out (with high confidence) classical local realistic theories
because classical strategies would be overwhelmingly unlikely
to perform flawlessly.
\end{itemize}

Of course, physical realizations of quantum winning strategies
for pseudo-telepathy games will \textit{never} be perfect.
Disavowing one hundred years of \mbox{quantum} mechanics on a single
failure---as we light-heartedly suggested above---would be foolish!
We~should be content as long as real-life experiments allow Alice
and Bob to win substantially more often than any classical players
could hope to do, \mbox{except} with overwhelmingly small probability.
We~expound on this issue in the next \mbox{section}.

But even an unfailing sequence of wins would not be entirely convincing
if the experiment is not performed with \textit{extreme} care.
There are many ways in which unscrupulous Alice and Bob---or~an
incredibly devious Nature out to fool us into believing in
quantum mechanics---could provide classical (local, realistic) players
with a winning strategy despite the mathematical proofs that this is
impossible!
Before they can be used to rule out classical theories
beyond any reasonable doubt,
pseudo-telepathy experiments must fulfil the following
conditions simultaneously.
\begin{itemize}
\item[$\Diamond$]
It must be physically impossible for Alice and Bob to communicate
between the time they receive the question and the time
they have to produce the answer.
No~amount of shielding can achieve this requirement.  For instance,
it~would be unscientific to claim it impossible for Alice to
have a means to modulate neutrinos and for Bob to detect them,
however unlikely.
The only known law of physics that
can be used to enforce this restriction is the fact that information cannot
be signalled faster than the speed of light. Hence, Alice and Bob must be
sufficiently far apart, their questions must be asked with near-perfect
\mbox{simultaneity}, and they must be requested to answer very quickly.
\item[$\Diamond$]
It must be impossible for Alice and Bob to know ahead of time what the
questions will be.
We~must first come to grip with the fact that there is
no way to enforce this condition if the universe is purely deterministic.
In~the view of Laplace, nothing prevents Alice and Bob from analysing the
entire state of the universe with enough precision to be able to predict
with certainty what questions Xavier and Yolande will ask.
This would allow Alice and Bob to conspire together and decide
on their answers before
they are separated, and Xavier and Yolande would be none the wiser.
Unfortunately, there is no resolution to this problem: no pseudo-telepathy
experiment will \emph{ever} convince a determinism-addict that classical
physics is wrong.
\item[$\Diamond$]
Following up on the previous point, the questions must be determined
at random by Xavier and Yolande \textit{after they have separated}
to go with Alice and Bob, respectively. Indeed, we said earlier that
``Xavier and Yolande are also allowed to discuss strategy in secret
of Alice and Bob''.  But no laws of physics can guarantee them
unconditional secrecy (not even quantum cryptography!)\ if they
choose their questions ahead of time.  Therefore, each round must
begin with Xavier and Yolande choosing their questions
purely at random, independently from one another.
The entire round, including registering Alice's and Bob's answers,
must take place in less time than it takes light to travel between Alice
and Bob.  This brings up a subtle point: given that Xavier and Yolande
cannot prepare their questions together, there can be no guarantee
that the promise will be fulfilled.
After they meet again, Xavier and Yolande will have to apply post-selection
to their data to keep only the answers to legitimate questions.
It~follows that it is desirable for pseudo-telepathy games
to have a high density of legitimate questions.
\item[$\Diamond$]
And finally, the last condition to rule out classical local realistic
theories is that Xavier and Yolande's post-selected data must show that
appropriate answers were provided on legitimate questions
\textit{significantly more often} by Alice and Bob than could
be expected by any possible classical process.
\end{itemize}

\subsection{Imperfect Detectors}\label{loopholes}

Several reasons could explain an \mbox{incorrect} answer
provided by the physical implementation of a mathematically-proven
quantum winning strategy:
(1)~quantum mechanics is incorrect,
(2)~Alice and Bob are unable to keep high-grade entanglement
for a macroscopic amount of time, (3)~the unitary quantum process to which
they subject their entanglement after receiving their questions
(see~all the following sections) is imperfect, and/or
(4)~the final measurement that should reveal the classical answer
goes wrong.

Here we concentrate on the last possibility.  We~ask how bad can the
detectors be to allow nevertheless for an outcome that would be
classically impossible (in~probability), provided everything else
is perfect in the experiment.  For this purpose, we must distinguish
between two different types of imperfections in detectors:
noise and inefficiency.
\begin{itemize}
\item[$\Diamond$] A \textit{noisy} detector may register the wrong result.
This causes Alice or Bob to output a bit that is the complement of what
she or he should have produced according to the protocol\,%
\footnote{\,In some pseudo-telepathy games, such as the
impossible colouring game of Section~\ref{section:impossiblecoulouring},
the expected output is \textit{not} a bit string.
The notion of noisy detector must be modified accordingly.}.
In~most pseudo-telepathy games, a single error of this sort will
result in an inappropriate answer to a legitimate question.
But~it is often the case that two wrongs make a right.
\item[$\Diamond$] An \textit{inefficient} detector may fail to register
an outcome at all.  When this happens, the qubit that was supposed to be
measured is usually lost forever, and there is nothing the unlucky
party can do to recover the information (without communicating with
the other party).  But at least, contrary to the case of noisy detectors,
that party knows that there is a problem.
\item[$\Diamond$] Of course, real detectors can be both noisy and
inefficient.
\end{itemize}
In each case, the relevant issue is to determine the level of imperfection
detectors can have, and still allow Alice and Bob to accomplish a task
that would exceed the capabilities of classical players.

\subsubsection{Noisy Detectors}

For simplicity, let us assume that all the detectors are \mbox{independent}
and that they all have the same error probability.
Let $p$ stand for the probability for each detector to perform
correctly, and therefore each bit of the answer is flipped
with probability \mbox{$1-p$}, independently from one another.
For any given pseudo-telepathy game, this entails a probability
for quantum players to provide inappropriate answers to legitimate
questions.
If that probability exceeds the error probability that can be
reached by some classical strategy, the quantum implementation
becomes worthless.  This gives rise to a threshold that depends
on the pseudo-telepathy game being played.

\newpage

\begin{defn}
Consider a pseudo-telepathy game~$G$.
We~define $p_{*}(G)$ as the maximum value of $p$ for which a
classical strategy can succeed as well as an implementation of the best
possible quantum strategy that has to deal with detectors that give the
correct answer with probability~$p$.
\end{defn}

\noindent
Unfortunately, the exact value of $p_{*}(G)$
is unknown for most games.

\subsubsection{Inefficient Detectors}

If a detector fails to respond to Alice or Bob,
one approach would be to ignore the problem and output a random bit.
This would be a way to turn inefficient
detectors into noisy~ones.  But a much better solution is to allow
the parties to admit ignorance.  When a party's detector fails to register
the result of a measurement, this party simply outputs~$\bot$.
In~this context, we must redefine slightly the notion of a game.
The output sets $A$ and $B$ are augmented with this new symbol~$\bot$.
We~say that the players \textit{win} if the question is not legitimate
or if their answer is appropriate, it's a \textit{draw} if at least one
player outputs~$\bot$, and they \textit{lose} otherwise.
The~key observation is that if detector inefficiency is the \textit{only}
imperfection in the quantum players' implementation, they will
either win or draw, but they will never lose a round.

Of course, to be fair, we must allow classical players to output $\bot$
as well. This makes it possible for classical players to have an
\textit{error-free} strategy, in which they never lose,
at the expense of occasional draws.
Implementations of quantum strategies are interesting only if
the occurrence of~$\bot$ is smaller than with the best
error-free classical strategy.
Whether or not this is possible depends on the efficiency $\eta$
of the detectors, which is assumed to be the same,
independently, for each detector.

\begin{defn}
Consider a pseudo-telepathy game~$G$.
We~define $\eta_{*}(G)$ as the maximum value of $\eta$ for which an
error-free classical strategy can win at least as often as an
implementation of the best quantum strategy that uses detectors
of efficiency~$\eta$.
\end{defn}

Parameter $\eta_{*}(G)$ has been analysed for more games than~$p_{*}(G)$
because it bears direct relevance to the widely studied
``detection loophole''~\cite{Massar02,MP03}, but nevertheless
its exact value is also mostly unknown.

\subsection{Outline of the Paper}
After this Introduction, the rest of the paper provides a systematic
survey of all the pseudo-telepathy games known to the authors.
We~assume that the reader is already familiar with the basic
principles of quantum information processing~\cite{Chuang}.
The games are presented in chronological order of discovery.
Each game is defined by the task that the players must accomplish
in order to win, then a quantum winning strategy is described, and
the impossibility of a classical winning strategy is discussed.
Additional information is provided for some games.
Before we conclude with open problems in Section~\ref{conclusions},
we have a quick look in Section~\ref{frontier} at the smallest possible
games that exhibit pseudo-telepathy.

\section{Impossible Colouring Games}
\label{section:impossiblecoulouring}

In the 1960s, Bell was not alone in proving that Einstein's realism
is incompatible with other well-accepted principles~\cite{Bell}.
Recall that Bell's argument was to reach a contradiction by accepting also
Einstein's nonlocality, as well as the predictions of quantum mechanics.
At~about the same time, Simon Kochen and Ernst Specker proved that realism
is also
incompatible with another principle: \textit{non-contextuality}~\cite{KS67}.
Briefly stated, non-contextuality is the principle according to which the
probability of a given outcome in a projective measurement
does not depend on the choice of the other orthogonal outcomes
used to define that measurement.
(To~be historically precise, it should be noted that Bell himself
published a similar idea~\cite{Bell66} before Kochen and
Specker~\cite{KS67}, \textit{but} Specker by himself presented
an earlier proof as early as 1960~\cite{Specker60}, and a similar
idea goes back to Gleason in 1957~\cite{Gleason57}.)

It has been argued that the Kochen-Specker approach is less useful than
Bell's because it in inherently counterfactual, and therefore it seems
impossible to subject it to actual experiments. However, when
Kochen-Specker is used together with entanglement,
Peter Heywood and Michael Redhead~\cite{HR82} realized that it can
be turned into a testable scenario.
A~little imagination suffices to turn their argument into a
pseudo-telepathy game, which is probably the first such game ever.
However, that paper was overlooked until recently, which explains
why the first pseudo-telepathy game is often credited to
Greenberger, Horne and Zeilinger~\cite{GHZ}
or to Mermin~\cite{MerminGHZ}---see Section~\ref{parity}.

\begin{thm}[Kochen-Specker Theorem]
 \label{KStheorem}
 There exists an explicit, finite set of vectors
 in $\mathbb{R}^3$
 that cannot be $\{0,1\}$-coloured so that both of
 the following conditions hold simultaneously:
 \begin{enumerate}
    \item For every orthogonal pair of vectors,
    \emph{at most} one of them is coloured~1.
    \item For every mutually orthogonal triple of vectors,
    \emph{at least} one of them---and therefore exactly one---is coloured~1.
 \end{enumerate}
 We say of any such set of vectors that is has the
 \emph{Kochen-Specker property}.
\end{thm}

Originally, Theorem~\ref{KStheorem} was proved using 117 vectors%
~\cite{KS67}, but this has been reduced to 31 (with 17 orthogonal
triples) by Conway and Kochen~\cite{Peres}. A~Kochen-Specker
construction, similar to that of Theorem~\ref{KStheorem}, can be
obtained in any dimension $d \geq 3$, either by a geometric
argument, or by extending a construction in dimension $d$ to
dimension $d+1$~\cite{Peres}. Following the approach of Richard
Cleve, Peter H{\o}yer, Benjamin Toner and John Watrous
\cite{CHTW04}, we~show below that any Kochen-Specker construction
in any dimension gives rise to a two-party pseudo-telepathy game.
Renato Renner and Stefan Wolf have proved a weak converse of this
result: Any two-party pseudo-telepathy game can be turned into a
Kochen-Specker construction~\cite{RW03}, but under two conditions:
Alice and Bob must share a maximally entangled state (of~any
dimension) and they are restricted to making projective
\mbox{measurements} only (not~POVMs) on their shared entanglement,
without the addition of an ancillary \mbox{system}.

\subsection{The Game}

Consider any dimension $d \ge 3$ and let~$V$
be a set of vectors in~$\mathbb{R}^d$ with
the Kochen-Specker property. In~the corresponding impossible colouring
pseudo-telepathy game, Alice receives either an orthogonal pair of vectors
$v_1, v_2$, or an orthogonal $d$-tuple of vectors $v_1, \ldots , v_d$. Bob
receives a single vector $v_\ell$.  All~these vectors are taken
from~$V$\!\@.
The promise states that $v_\ell$ is
one of Alice's vectors.

The challenge that Alice and Bob face is that they must $\{0,1\}$-colour
the vectors they have received so that (1)~two orthogonal vectors are not
both coloured~1, (2)~exactly one of Alice's vectors is coloured~1
if she was given $d$ vectors, and (3)~Alice and Bob assign the same
colour to vector~$v_\ell$. Note that it is not necessary for Alice
to output the colour of all her vectors because she must satisfy
conditions (1) and~(2). In~case she is given two vectors as input,
she outputs a trit that indicates whether $v_1$ is coloured~1,
$v_2$ is coloured~1, or neither; in case she is given $d$ vectors,
she outputs the \mbox{$a \in \{1,2, \ldots , d \}$} so that
she assigned colour~1 to~$v_a$.  As~for Bob, his one-bit output $b$ is
simply the colour he assigns to vector~$v_\ell$.
In~any case, the
winning condition is that Alice and Bob must assign the same colour
to~$v_\ell$.

\subsection{Quantum Winning Strategy}
The players' strategy is to share entangled state
$\frac{1}{\sqrt{d}} \sum_{j=0}^{d-1} \ket{j}\ket{j}$. After
receiving their input, Alice and Bob do the \mbox{following}:

\begin{enumerate}
\item If Alice was given only two vectors $v_1$ and $v_2$, she chooses
\mbox{$d-2$} additional vectors $v_3$, \ldots, $v_d$,
not necessarily taken from~$V$\!, so that $v_1$, \ldots, $v_d$
forms an orthogonal $d$-tuple.
\item Alice performs a measurement on her share of the entangled state
in basis \mbox{$B_a=\{\ket{v_1}, \ldots, \ket{v_d}\}$} of~$\mathbb{R}^d$
(after normalization if necessary).
Let~$k$ be the result of her measurement.  As~explained above,
she produces the appropriate output~$a$, corresponding to assigning
colour~1 to vector~$v_k$.
\item Bob chooses \mbox{$d-1$} additional vectors
$w_1$, \ldots, $w_{d-1}$, not necessarily taken from~$V$\!,
so that $v_\ell$, $w_1$, \ldots, $w_{d-1}$ forms an orthogonal $d$-tuple.
He~performs a measurement on his share of the entangled state in \mbox{basis}
\mbox{$B_b=\{\ket{v_\ell}, \ket{w_1}, \ldots , \ket{w_{d-1}}\}$}
of~$\mathbb{R}^d$ (after normalization if necessary).
He~outputs $b=1$ if the outcome is $v_\ell$; he~outputs~$b=0$ otherwise.
\end{enumerate}

To show that this quantum strategy works, consider the probability that
Alice measures $v_k$ $(k \in \{1,2, \ldots ,d\})$ and Bob measures~$v_\ell$.
Since the bases $B_a$ and $B_b$ have real coefficients,
\begin{equation}
\sum_{j=0}^{d-1} \bra{j}v_k \rangle \bra{j}v_\ell \rangle =
\bra{v_k}v_\ell \rangle
\end{equation}
and so this probability is
\begin{align}
\left|\frac{1}{\sqrt{d}}
\sum_{j=0}^{d-1} \bra{j}\bra{j} v_k \rangle \ket{v_\ell}\right|^2 &
=\left|\frac{1}{\sqrt{d}} \sum_{j=0}^{d-1} \bra{j}v_k \rangle
\bra{j}v_\ell \rangle \right|^2 \\[2ex]
&= \left| \frac{1}{\sqrt{d}}\bra{v_k}v_\ell \rangle \right|^2 \\[2ex]
&= \begin{cases}
\frac{1}{d}, &k=\ell \\ 0, &k \neq \ell
\end{cases}
\end{align}

It follows that Bob measures~$v_\ell$ if and only if Alice
measures~$v_\ell$ as well.
This proves that Alice and Bob will assign the same colour to the vector
they have received in common, and therefore their answer will be
appropriate.

\subsection{Classical Players}
Any classical deterministic winning
strategy would correspond precisely to a $\{0,1\}$-colouring
of the vectors in~$V$ that can satisfy simultaneously properties
(1) and (2) in the statement of the Kochen-Specker theorem.
But~$V$ was chosen to have the Kochen-Specker property,
which means that such a colouring cannot exist.
It~follows that
\mbox{$\widetilde{\omega}(G) < 1$}, and by
Theorem~\ref{framework:probabilistic},
\mbox{$\omega(G) < 1$} as well.
Therefore, no classical winning strategy can exist for this game.
The exact success proportion and probability are not known precisely,
but of course they
depend on the the specific Kochen-Specker construction that is used.
We~conjecture that $\widetilde{\omega}(G)$ and $\omega(G)$
cannot be made much smaller than~1 with this type of construction.

\subsection{Special Case of the Impossible Colouring Game}
 We have presented a family of pseudo-telepathy games based on
the Kochen-Specker theorem.  It is interesting to mention the
particular case where $d=3$.
For the quantum strategy, Alice and Bob share an
entangled \emph{qutrit} pair
\mbox{\smash{$\ket{\psi}= \frac{1}{\sqrt{3}}\ket{00}
+ \frac{1}{\sqrt{3}}\ket{11}+\frac{1}{\sqrt{3}}\ket{22}$}}.
This entangled
state of dimension 9 is the smallest possible state that can be used
for any two-player pseudo-telepathy game~\cite{BMT04}---more on this
in Section~\ref{frontier}.

Independently of the general approach that we described above,
similar pseudo-telepathy games have been obtained for a specific
4-dimensional construction in~\cite{ZP93}, and generalized to any
dimension in~\cite{Aravind2}. See~also~\cite{JA99}.

\section{Parity Games}\label{parity}\label{section:paritygames}
This is a family of games for $n$
players, $n\geq 3$, with the property that the player's outputs
are single bits, and the winning condition depends on their parity.

\subsection{The Game}
The task that the $n$ players face is the following:  Each player
$i$ receives as input a bit-string $x_i \in \{0,1\}^\ell$, which
is also interpreted as an integer in binary, with the promise
that $\sum_{i=1}^n x_i$ is divisible by $2^{\ell}$. The players
must each output a single bit $a_i$ and the winning condition is:
\begin{equation}
 \sum_{i=1}^n a_i  \equiv
\frac{\sum_{i=1}^n x_i}{2^{\ell}} \pmod 2
\end{equation}

When restricted to parameters $n=3$ and $\ell=1$, this is what is
known as the Mermin-GHZ game. It~was originally
presented as a four-player game~\cite{GHZ}
and later recast with three players~\cite{GHSZ}.
The game was
greatly popularized by Mermin~\cite{MerminGHZ,Mermin:PhysToday}.

When restricted to $\ell=1$, but with any number $n \geq 3$ of players,
this is Mermin's \emph{parity game}, introduced in~\cite{Mermin90}
and recast from a computer science point of view in~\cite{BBT03,BBT04}.

 Finally, by setting $\ell=\ceil{\,\lg n}-1$, still with any number
 $n \geq 3$ of players,
  we get the \emph{extended parity game}, proposed by
 Harry Buhrman, Peter H{\o}yer, Serge Massar
  and Hein R{\"ohrig}~\cite{BHMR03}.
(We~use the symbol ``$\lg$'' to denote the base-two logarithm.)

\subsection{Quantum Winning Strategy}
The players' strategy for any game in this family is to share entangled
state \smash{$\phiplusn$}. After receiving his input~$x_i$,
each player $i$ does the following:
\begin{enumerate}
\item  \label{stepBHMR2}
 apply to his share of the entangled state the unitary
transformation  $S$ defined by
\begin{align}
&\ket{0} \mapsto \ket{0}\\
&\ket{1} \mapsto e^{\pi \ii x_i / 2^\ell}\ket{1}
\end{align}
where we use a dotless $\ii$
to denote $\sqrt{-1}$ in order to distinguish it from index $i$,
which is used to identify a player
\item \label{stepBHMR3} apply the Walsh--Hadamard transform, $H$,
defined as usual by
\begin{align}
&\ket{0} \mapsto
{\textstyle \frac{1}{\sqrt{2}}\ket{0}+\frac{1}{\sqrt{2}}\ket{1}} \\
&\ket{1} \mapsto
{\textstyle \frac{1}{\sqrt{2}}\ket{0}-\frac{1}{\sqrt{2}}\ket{1}}
\end{align}
 \item \label{stepBHMRmeasure}measure the qubit in the
 computational basis to obtain $a_i$
 \item \label{stepBHMRoutput}output $a_i$
\end{enumerate}

The resulting state after step~\ref{stepBHMR2} is
\begin{equation}
{\textstyle \frac{1}{\sqrt{2}}} \left( \ket{0^n} +
   e^{\pi \ii \frac{\sum x_i}{2^\ell}}\ket{1^n} \right)\\[2ex]
~=~
\begin{cases} \phiplusn~~ &
  \text{if~} \frac{\sum x_i}{2^\ell} \text{~is even} \\[2ex]
\phiminusn &
  \text{if~} \frac{\sum x_i}{2^\ell} \text{~is odd}
\end{cases}
\end{equation}

\noindent
We know by the promise that $\frac{\sum x_i}{2^\ell}$ is an
integer.  Let $\Delta(z)$ denote the number of 1s in binary string~$z$
(the~\emph{Hamming weight} of~$z$).
It~is an easy exercise to show that the
resulting state after step~\ref{stepBHMR3} is:
\begin{align}
\begin{cases} \displaystyle
\frac{1}{\sqrt{2^{n-1}}}\sum_{\Delta(a) \text{ even}} \ket{a}~~ &
\displaystyle \text{if~} \frac{\sum x_i}{2^\ell} \text{~is even} \\[3ex]
\displaystyle \frac{1}{\sqrt{2^{n-1}}}\sum_{\Delta(a) \text{ odd}} \ket{a} &
\displaystyle \text{if~} \frac{\sum x_i}{2^\ell} \text{~is odd}
\end{cases}
\end{align}
Therefore, after the measurement of step~\ref{stepBHMRmeasure}, the
output of step~\ref{stepBHMRoutput} will
 satisfy:
\begin{equation}
 \sum_{i=1}^n a_i  \equiv
\frac{\sum_i^n x_i}{2^\ell} \pmod 2
\end{equation}
so the players always win.

\subsection{Classical Players}\label{no-classical-parity}

Any winning strategy for the extended
parity game yields a winning strategy for the parity game.
Similarly, a winning strategy for the parity game yields a winning
strategy for the Mermin-GHZ game.  So, by showing that there is no
classical winning strategy for the Mermin-GHZ game, we will have
established the same result for all three games.

In the Mermin-GHZ game, consider a deterministic strategy in which
$a_x$ is Alice's output on input $x$, and $b_y$, $c_z$ are defined
similarly for Bob and Charlie.  If Alice, Bob and Charlie receive
$x=0$, $y=0$, $z=0$ as inputs, for instance, it follows from the winning
condition  that the sum of their outputs $a_0$, $b_0$ and $c_0$
must be even.  This gives the first of the four constraints that
correspond to the four legitimate inputs $000, 011, 101$ and
$110$.  These constraints are summarized below,
where the sums are taken modulo 2.
\begin{equation}
 \label{GHZclassical}
 \begin{array}{ccccc}
&a_0 + b_0 + c_0  \equiv 0  \\
   &a_0 + b_1 + c_1  \equiv 1 \\
   &a_1 +   b_0 + c_1 \equiv 1 \\
 & a_1 + b_1 +  c_0 \equiv 1 \\
 \end{array}
\end{equation}
If we add the four equations, we obtain a contradiction, since the
sum on the left-hand side is even, and the sum on the right-hand
side is odd.  This proves that
 there is no
classical deterministic winning strategy.  It follows by Theorem%
~\ref{framework:probabilistic} that there is no classical winning strategy.
However, it is easy to invent a strategy that wins with
probability~\pbfrac{3}{4}.
It~follows from Lemma~\ref{lemma:proportion} that
\mbox{$\widetilde{\omega}(G)=\omega(G)=\pbfrac{3}{4}$},
where $G$ stands for the Mermin-GHZ game.

Let $G_n$ be Mermin's parity game with $n$ players.  It is shown in%
~\cite{BBT03} that \mbox{$\widetilde{\omega}(G_n) = \optprob$}.
By~applying Theorem~\ref{framework:probabilistic}, it follows that
\mbox{$\omega(G_n) \leq \optprob$}.
But in fact, this upper-bound is tight~\cite{BBT04,AnneMasters}:
\mbox{$\omega(G_n) = \optprob$}.

For the parity game, we also know exact values of  $p_*$
   and $\eta_*$~\cite{BBT04,AnneMasters}:
\begin{align}
& p_*(G_n) =\begin{cases} \frac{1}{2} + 2^\frac{2-3n}{2n}~~ &
\text{if~} n \text{~is even} \\[2ex]
\frac{1}{2} + 2^\frac{1-3n}{2n} &
\text{if~} n \text{~is odd} \end{cases} \\[2ex]
& \eta_*(G_n) = {\textstyle \frac{1}{2}} \sqrt[n]{4}
\end{align}

Since
\smash{$ \lim_{n \rightarrow \infty}
p_*(G_n) = \frac{1}{2} + \frac{\sqrt{2}}{4}
= \cos^2 \pi/8 \approx 85 \%$}, we
conclude that if we use detectors that provide the correct
answer with a probability greater than that,
everything else being perfect, and if $n$ is sufficiently
large, we can use game $G_n$ to observe a phenomenon that
would be statistically impossible to explain in a classical world.
Similarly, since
$\lim_{n \rightarrow \infty} \eta_*(G_n) = \pbfrac{1}{2}$,
the same conclusion applies provided we have detectors that are
better than 50\% efficient.

Let $G'_n$  be the \textit{extended} parity game with $n$ players.
It~is shown in~\cite{BHMR03} that
\mbox{$\eta_*(G'_n) \leq \neopbfrac{8}{n}$} for all \mbox{$n \ge 3$}.
Since  \mbox{$\lim_{n \rightarrow \infty} \eta_*(G'_n) = 0$},
loophole-free experiments of nonlocality are possible with
arbitrarily inefficient detectors!
(Of~course, as $n$ becomes larger, it becomes increasingly difficult
to have ``everything else being perfect''.)
The~value of $p_*(G'_n)$ is not known.

\section{{D}eutsch-{J}ozsa Games}\label{DJgames}

This family of two-player pseudo-telepathy games, based on the
Deutsch-Jozsa problem~\cite{DJ92},
 was first presented in~\cite{DistDeutsch-Jozsa}.
These were the first games \textit{explicitly} \mbox{described} in terms
of pseudo-telepathy.

\subsection{The Game}

Alice and Bob face the following task:
given bit strings $x$ and $y$ as input,
they are requested to output bit strings $a$ and $b$, respectively,
so that $a=b$ if and only if $x=y$.
To~prevent the trivial solution $a \leftarrow x$ and $b \leftarrow y$,
the outputs must be exponentially shorter than the inputs.
More precisely, $x$ and $y$ are strings of length~\mbox{$n=2^m$},
for some parameter~$m$, but $a$ and $b$ must be strings of length~$m$.
As~such, this problem cannot be solved even with the help
of entanglement, but it becomes a pseudo-telepathy game if we add
the promise that either $x$ and $y$
are identical or that they differ in exactly half of the bit positions.

\subsection{Quantum Winning Strategy}

The players' strategy for any game in this family is to share entangled
state
\mbox{$\frac{1}{\sqrt{n}} \sum_{j=0}^{n-1} \ket{j}\ket{j}$}.
Note that this means that Alice and Bob each have an $m$-qubit register.
After receiving their inputs \mbox{$x = x_0 x_1 \cdots x_{n-1}$} and
\mbox{$y = y_0 y_1 \cdots y_{n-1}$}
(it~is more convenient to start the indices at~0),
Alice and Bob do the following:
\begin{enumerate}
\item Alice applies to her quantum register the unitary transformation
that maps
\begin{equation}
\ket{j} \mapsto (-1)^{x_j} \ket{j}
\end{equation}
for all $j$ between 0 and~\mbox{$n-1$};
Bob does the same, but with $y_j$ instead of~$x_j$
\item
Alice and Bob apply the Walsh--Hadamard transform $H^{\otimes m}$
to their registers
\item
Alice and Bob measure their registers in the computational basis.
The \mbox{resulting} classical strings, $a$ and~$b$, are their final output
\end{enumerate}
It is straightforward to verify that this quantum strategy wins
the game with \mbox{certainty}.
Please consult~\cite{DistDeutsch-Jozsa} for details.

\subsection{Classical Players}

It follows immediately from~\cite{BCW} that classical players
need to communicate at least $c\,2^m$ bits, for some appropriate positive
constant~$c$ and all sufficiently large~$m$, in order to
win the Deutsch-Jozsa game with parameter~$m$.
Therefore, classical players cannot have a winning strategy
if they are unable to communicate, again provided $m$ is large enough.
However, this asymptotic result does not help if we want
to know \textit{which} values of $m$ yield a pseudo-telepathy game.

It is easy to design a classical winning strategy for $m=1$
and $m=2$.  It~is more challenging to design one
when $m=3$, but this is done in~\cite{GW02}.
For the case $m=4$, it was shown
in~\cite{GTW03}, by using an argument based on graph theory as
well as a computer-assisted case analysis, that there is no
classical winning strategy.  However, the proof of this result does not
carry over to the case of larger values of~$m$, so it still remains to
determine explicitly for what other values of~$m$ pseudo-telepathy occurs.

\section{The Magic Square Game}\label{magic}
The magic square game is a two-player pseudo-telepathy game that
was presented by Padmanabhan Aravind~\cite{Aravind1,AravindX}, who
built on work by Mermin~\cite{MerminMagicSquare}.
The most interesting feature of this game is that it
is extremely easy to show that there cannot be a classical winning
strategy (see~Section~\ref{no-classical-magic}).
It~follows that a successful implementation of the quantum
winning strategy (see~Section~\ref{quantum-magic})
would convince any observer that something classically
impossible is happening, \textit{with no need for the observer to
understand why the quantum strategy works}.

\subsection{The Game}

A \emph{magic square} is a \mbox{$3 \times 3$} matrix
whose entries are in \mbox{$\{0,1\}$}, with the
property that the sum of each row is even and the sum of each
column is odd.  Such a square is magic because it cannot exist!
Indeed, suppose we calculate the parity of the nine entries.  According to
the rows, the parity is even, yet according to the columns, the
parity is odd. This is obviously impossible.

The task that the players face while playing the game is the
following: Alice is asked to give the entries of a row $x \in
\{1,2,3\}$ and Bob is asked to give the entries of a column $y \in
\{1,2,3\}$. The winning condition is that the parity of the row
must be even, the parity of the column must be odd, and the
intersection of the given row and column must agree.  It is an
interesting feature of this game that it does not require a promise:
all nine possible questions are legitimate.

\subsection{Quantum Winning Strategy}\label{quantum-magic}

The quantum winning strategy for the magic square game is not
as simple as the classical impossibility proof.
First, Alice and Bob share the entangled state
\begin{equation}
\textstyle \ket{\psi} = \frac{1}{2}  \ket{0011} -\frac{1}{2}\ket{0110}-
\frac{1}{2}\ket{1001} + \frac{1}{2} \ket{1100}
\end{equation}
The first two qubits belong to Alice and the last two to Bob.
Upon receiving their inputs $x$ and $y$, Alice and Bob
apply unitary transformations $A_x$ and $B_y$, respectively,
according to the following matrices.

\begin{equation}
A_1=\frac{1}{\sqrt{2}}\left[ \begin{smallmatrix}
     \ii & \Z 0 & \Z 0 & \Z 1 \\
     0 &   -\ii & \Z 1 & \Z 0 \\
     0 & \Z \ii & \Z 1 & \Z 0 \\
     1 & \Z 0 & \Z 0 & \Z \ii \\
  \end{smallmatrix} \right],~~
A_2=\frac{1}{2}\left[ \begin{smallmatrix}
  \Z \ii & \Z 1 & \Z 1 & \Z \ii \\
    -\ii & \Z 1 &   -1 & \Z \ii \\
   \Z \ii& \Z 1 &   -1 &   -\ii \\
    -\ii & \Z 1 & \Z 1  &   -\ii\\
  \end{smallmatrix} \right],~~
A_3=\frac{1}{2}\left[ \begin{smallmatrix}
    -1 &   -1 &   -1 & \Z 1 \\
  \Z 1 & \Z 1 &   -1 & \Z 1 \\
  \Z 1 &   -1 & \Z 1 & \Z 1 \\
  \Z 1 &   -1 &   -1 &   -1 \\
  \end{smallmatrix} \right]
\end{equation}

\begin{equation}
B_1=\frac{1}{2}\left[ \begin{smallmatrix}
 \Z  \ii &   -\ii & \Z 1 & \Z 1 \\
    -\ii &   -\ii & \Z 1 &   -1 \\
 \Z  1 & \Z 1 &   -\ii & \Z \ii \\
    -\ii & \Z \ii & \Z 1 & \Z 1 \\
  \end{smallmatrix} \right],~~
B_2=\frac{1}{2}\left[ \begin{smallmatrix}
    -1 & \Z \ii & \Z 1 & \Z \ii \\
  \Z 1 & \Z \ii & \Z 1 &   -\ii \\
  \Z 1 &   -\ii & \Z 1 & \Z \ii \\
    -1 &   -\ii & \Z 1 &   -\ii \\
  \end{smallmatrix} \right],~~
B_3=\frac{1}{\sqrt{2}}\left[ \begin{smallmatrix}
  \Z 1 & \Z 0 & \Z 0 & \Z 1 \\
    -1 & \Z 0 & \Z 0 & \Z 1 \\
  \Z 0 & \Z 1 & \Z 1 & \Z 0 \\
  \Z 0 & \Z 1 &   -1 & \Z 0 \\
  \end{smallmatrix} \right]
\end{equation}

Then, Alice and Bob measure their qubits in the computational basis.
This provides 2 bits to each player, which are the first 2 bits of
their respective output $a$ and~$b$.
Finally, Alice and Bob determine their third output bit
from the first two so that their parity condition is satisfied.

Consider for example inputs \mbox{$x=2$} and~\mbox{$y=3$}.
After Alice and Bob apply $A_2$ and~$B_3$, respectively, the state
evolves to
\begin{align}
(A_2 \otimes B_3)\ket{\psi} =
{\textstyle \frac{1}{2\sqrt{2}}}
   \bigl[ & \ket{0000} - \ket{0010} - \ket{0101} + \ket{0111}  \\
 \mbox{}+ & \ket{1001} + \ket{1011} - \ket{1100} - \ket{1110} \bigr]
\end{align}
After measurement, Alice and Bob could obtain 10 and~01, for instance.
In~that case, Alice would complete with bit~1 so that her output
\mbox{$a=101$} has even parity and Bob would complete with bit~0
so that his output \mbox{$b=010$} has odd parity.
Xavier and Yolande will be satisfied with the answer since
both Alice and Bob agree that the third entry of the second row
is indeed the same as the second entry of the third column:
\mbox{$a_3 = b_2 = 1$}.
It~is easy to check that the seven other possible answers that Alice and
Bob could have given on this example are all appropriate.
The~verification that this quantum strategy wins also on the other eight
possible questions is tedious but straightforward.

\subsection{Classical Players}\label{no-classical-magic}

The simple proof that a winning classical strategy
cannot exist goes as follows.
A~deterministic classical strategy would have to assign
definite binary values to each of the
nine entries of the magic square, which is impossible.
Therefore, there can be no deterministic classical winning
strategy.  It follows from Theorem~\ref{framework:probabilistic} that
there is no classical winning strategy, even probabilistic.
Using Lemma~\ref{lemma:proportion}, it is also straightforward
to show that \mbox{$\widetilde{\omega}(G)=\omega(G)=\pbfrac{8}{9}$},
where $G$ stands for the magic square game.

\subsection{Related Work}
There are other pseudo-telepathy games that are related to the
magic square game. Ad\'an Cabello's game~\cite{Cabello,cabello2}
does not resembles the magic square game on first approach.
However, closer analysis reveals that the two games are
totally equivalent!

Also, Aravind has generalized his own magic square idea~\cite{Aravind1}
to a two-player pseudo-telepathy
game in which the players share $n$ Bell states, $n$ being an arbitrary
odd number larger than~1.

\section{Matching Games}\label{matching}
This family of two-player games is a relatively recent
development.  It is based on an observation made by Harry Buhrman
in response to a talk given by Iordanis Kerenidis at the 2004~Workshop
on Quantum Information Processing~\cite{BK04}, and on~\cite{BJK04}.

\begin{defn}
 A \emph{perfect matching}
$M$ on $\{0,1, \ldots ,m-1\}$, where $m$ is even, is a partition of
\mbox{$\{0,1, \ldots m-1 \}$} into $\frac{m}{2}$ sets, each of size 2.
We define $M_m$ as the set of all perfect matchings on
$\{0,1, \ldots ,m-1\}$.
\end{defn}

\subsection{The Game}

In the matching game, Alice receives an $m$-bit string
\mbox{$x=x_0 x_1 \cdots x_{m-1}$} as input
(again, it is more convenient to start the indices at~0),
and Bob receives a perfect matching \mbox{$M \in M_m$}. The task
that the players face is for Alice to output a string
\mbox{$a \in \{0,1\}^{\ceil{\,\lg m}}$}, and Bob to output a pair
\mbox{$\{\alpha,\beta\} \in M$} as well as a  string
\mbox{$b \in \{0,1\}^{\ceil{\,\lg m}}$} such that
\begin{equation}
x_\alpha \oplus x_\beta = \left(\alpha \oplus \beta \right) \cdot
\left(a \oplus b\right)
\end{equation}
where \mbox{$a \oplus b$} denotes the bitwise exclusive-or of $a$ and $b$,
and
\mbox{$u \cdot v = \bigoplus_i (u_i \wedge v_i$)}
\mbox{denotes} the parity of the bitwise ``\textsf{and}'' of $u$ and~$v$.

This may seem at first to be a very bizarre game, but it is
currently the only \textit{scalable} game known---it~depends on a
parameter~$m$---among those that do not require a promise.

\subsection{Quantum Winning Strategy}

The quantum players' strategy is to share entangled state
$\frac{1}{\sqrt{m}} \sum_{j=0}^{m-1}\ket{j}\ket{j}$. After
Alice receives her input \mbox{$x=x_0 x_1 \cdots x_{m-1}$}
and Bob his input \mbox{$M \in M_m$}, the players do the following:
\begin{enumerate}

\item \label{BK1}
Alice applies to her quantum register the unitary transformation
that maps
\begin{equation}
\ket{j} \mapsto (-1)^{x_j} \ket{j}
\end{equation}
for all $j$ between 0 and~\mbox{$m-1$}

\item \label{BKBOBmeasure}
Bob performs a projective partial measurement onto subspaces of dimension~2.
Each subspace of the measurement is spanned by
vectors $\ket{k}$ and $\ket{\ell}$, where \mbox{$\{k,\ell\} \in M$}.
Bob outputs the classical outcome of this measurement,
which is a pair \mbox{$\{\alpha, \beta\} \in M$}.
In~addition to producing a classical outcome, the measurement
causes the quantum state shared with Alice to collapse to
\mbox{$\frac{1}{\sqrt{2}}(-1)^{x_\alpha} \ket{\alpha \alpha} +
\frac{1}{\sqrt{2}}(-1)^{x_\beta} \ket{\beta \beta}$}

\item \label{BKH}Both Alice and Bob perform the Walsh--Hadamard transform
$H^{\otimes \ceil{\,\lg m}}$

\item \label{BKAliceMea} Alice measures in the computational basis and
outputs $a$, the result of her measurement

\item \label{BKBobMea}Bob measures in the computational basis  and
outputs $b$, the result of his measurement

\end{enumerate}

\noindent
Details of why this works are left as an exercise.

\subsection{Classical Players}

By reducing the \emph{Hidden Matching Problem}~\cite{BHMR03}
to the matching game, it follows that there is no classical winning
strategy for the latter, provided $m$ is chosen large enough.
Once again, this does not tell us exactly which values of
$m$ yield a pseudo-telepathy game. For now, we are only able to
say that there is a trivial classical winning strategy for~\mbox{$m=2$}.

\section{Other Games}

We report briefly on two additional families of pseudo-telepathy games.

The first one is due to David DiVincenzo and Asher Peres%
~\cite{DP97}.  By exploiting properties of quantum code words for
3, 5, 7 and 9 qubits codes, they have given an elegant algebraic
way of generating 3, 5, 7 and 9 player pseudo-telepathy games,
where the shared entangled state is a quantum code word. To the
best of our knowledge, these games are no harder to win for
classical players than the parity games of
Section~\ref{section:paritygames} for the same number of players.

The second game is due to Michel Boyer~\cite{B04}.  It is a
multi-party pseudo-telepathy game for \mbox{$n \geq 3$} players.
Let~$M$ be any even number.
The task that the players face is the following:  Each
player $i$ receives as input an arbitrary integer $x_i$ and
must output an integer $a_i$ between 0 and~\mbox{$M-1$}.
The promise is that $\sum x_i$ is even and
the winning condition is that
\begin{equation}
\sum_{i=1}^n a_i \equiv \frac{\sum_{i=1}^n x_i}{2} \pmod M
\end{equation}
This game is equivalent to a variant in which each input is
a single bit.  One interest of this game is that the quantum winning
strategy uses a quantum Fourier transform modulo~$M$.
It~may be more convenient to take $M$ a power of~2.

\section{The Inner Frontier}\label{frontier}

Throughout this survey, we have presented several
pseudo-telepathy games, large and small
%CAREFUL: there should NOT be a comma after "small" above!!!
according to a
variety of metrics such as the cardinality of the input sets,
of the output sets, and the dimensionality of the entangled
state that must be shared at the outset of the game.
It~it interesting to investigate what are the \textit{smallest}
possible games that exhibit pseudo-telepathy.

Consider two-party pseudo-telepathy games.
It~is known that no such game can exist if Alice and Bob
are requested to output a single bit each~\cite{CHTW04}.
It~follows that the impossible colouring game
(Section~\ref{section:impossiblecoulouring})
with \mbox{$m=3$}
is minimal in terms of the output size since Alice outputs a trit
and Bob outputs a bit.  Another way of seeing this inner frontier
is that the Mermin-GHZ game
(Section~\ref{section:paritygames})
is also minimal since it exhibits
pseudo-telepathy with a single bit of output per player\ldots{}
but it involves \textit{three} players.

Similarly, two-party pseudo-telepathy games cannot exist
if the entangled state that the quantum players share is of dimension
\mbox{$2 \times 3$} or smaller~\cite{BMT04}.
It~follows that the impossible colouring game with \mbox{$m=3$}
is again minimal since it makes do with an entangled state
of dimension \mbox{$3 \times 3$}.
Or~one could say that it's the Mermin-GHZ game that is again minimal
since it exhibits pseudo-telepathy with a single qubit of entanglement
in the hands of each of the three players.
In~terms of the total dimensionality, it is the Mermin-GHZ game that wins
the minimalist Grand Prize since \mbox{$8<9$}, where
\mbox{$8 = 2 \times 2 \times 2$} and~\mbox{$9 = 3 \times 3$}.

\section{Conclusions and Open Problems}\label{conclusions}

Einstein, Bohr, Bell, Kochen and Specker
were all concerned with hidden variables
(``elements of reality'').
Pseudo-telepathy also deals with this issue:
it provides alternate versions of Bell's argument
against local realistic theories. But it's more than just that.
Pseudo-telepathy games often provide a more concise and convincing
argument than those along the lines of Bell.  They may also
prove useful in devising loophole-free experimental tests to rule
out local realistic descriptions of the physical world.

The seeds of
 pseudo-telepathy were sowed more  than twenty years ago
 by  Heywood and Redhead (Section~\ref{section:impossiblecoulouring}).
  Others followed.
 By~setting these experiments in the framework of pseudo-telepathy, we
 are  better able to grasp the essence of the arguments,
 compare them,
  and improve on previous results.  For example, it is only recently,
 and in
 the context of pseudo-telepathy, that it was realized that the
 two earliest pseudo-telepathy games were minimal in term
 of the number of possible outputs as well as the dimensionality of the
 required entanglement (Section~\ref{frontier}).

Several interesting questions are still open. We have seen in
Section~\ref{loopholes}  that noisy and inefficient
detectors will unavoidably render the quantum players imperfect.
For the parity games of Section~\ref{section:paritygames}, we
have given exact values or bounds on the efficiencies that we can
tolerate. In order to devise experiments that are less prone to
these types of errors, it would be important to have an analysis
for each game $G$ that  provides exact values of $p_*(G)$ and
$\eta_*(G)$.  Also, imperfections can arise from the entangled
state and the unitary evolution of the system. An analysis of
these types of noise, as well as how the different types interact,
would give us  the tools necessary  to devise better experiments.

Of course, it would be interesting to find new pseudo-telepathy
games or families of games.  It~would be equally interesting to
show how they relate to one another.  For example, the three games
of Section~\ref{section:paritygames} can be parametrized so that
they all fall into the same general description.  The magic square
game (Section~\ref{magic}) is equivalent to Cabello's game.  Which
other games, old or new, are similar?  How do they differ?

The matching games of Section~\ref{matching} come
from a one-way communication complexity problem. Is it possible
to find other links between one-way communication and
pseudo-telepathy? Still concerning the matching
game $G_m$ with parameter $m$, it would be interesting to find
values for $\widetilde{\omega}(G_m)$ and $\omega(G_m)$ for even $m
\geq 4$. There are other games, such as the impossible colouring
games of Section~\ref{section:impossiblecoulouring},
the extended parity games of Section~\ref{section:paritygames}
and the Deutsch-Jozsa games of Section~\ref{DJgames}, for which, in general,
precise values of $\widetilde{\omega}(G)$ and $\omega(G)$, as a
function of the size of the game, are still unknown.

In Section~\ref{frontier}, we gave bounds on the dimension
of the shared entangled state and on the cardinality of the output
sets in order for a two-player pseudo-telepathy game
to exist.  Can we find similar bounds on the cardinality of the
\textit{input} sets?  Is~the magic square game (Section~\ref{magic})
optimal in this respect?

Finally, it is not enough to find games $G$ that have a small
classical success proportion $\widetilde{\omega}(G)$ or
probability $\omega(G)$, or a good tolerance to detector noise
$p_*(G)$ or inefficiencies $\eta_*(G)$.  We must strive to find
such games that are feasible experimentally.  The most simplistic
interpretation of this condition is  that the classical success
proportion or probability should be small, or the tolerance to detector
noise or inefficiencies should be high, even for \emph{small sized}
entangled quantum systems.  We would like to know bounds on what
is achievable for pseudo-telepathy games, as well as games that
reach these bounds. Failing that, it might be possible to find new
pseudo-telepathy games that improve on any of the best known
% CAREFUL... No hyphen above in "best known" !
values for the above parameters, as a function of the size of the
entangled quantum state.

\section*{Acknowledgements}

We are grateful to Michel Boyer, Richard Cleve, Serge Massar and
Andr\'e M\'ethot  for their insightful comments.
We are also grateful to Harry Buhrman and Iordanis Kerenidis
for sharing their thoughts on the matching game.
We are especially
grateful to Asher Peres for having been an ongoing inspiration for
our study of quantum information in general and nonlocality in particular,
as well as for his excellent book~\cite{Peres}.

% ----------------------------------------------------------------

\end{document}